\documentclass[10pt,twocolumn,twoside]{IEEEtran}

\usepackage{amsmath,epsfig,graphicx,color,multirow,subfigure}

\begin{document}

\title{Cooperative Game-Theoretic Approach to Spectrum Sharing in Cognitive Radios}

\author{Jayaprakash~Rajasekharan,~\IEEEmembership{Student,~IEEE,} Jan~Eriksson,~\IEEEmembership{Member,~IEEE,} and~Visa~Koivunen,~\IEEEmembership{Fellow,~IEEE}

\thanks{J. Rajasekharan, J. Eriksson and V. Koivunen are with SMARAD CoE, Department of Signal Processing, Aalto University, P.O.Box 13000, FI-00076, Espoo, Finland. e-mail: \{jrajasek,jamaer,visa\}@signal.hut.fi.}

\thanks{Some preliminary ideas and results were published in Asilomar 2010 and IEEE PIMRC 2011 conferences.}}

\maketitle

\begin{abstract}

In this paper, a novel framework for normative modeling of the spectrum sensing and sharing problem in cognitive radios (CRs) as a transferable utility (TU) cooperative game is proposed. Secondary users (SUs) jointly sense the spectrum and cooperatively detect the primary user (PU) activity for identifying and accessing unoccupied spectrum bands. The games are designed to be balanced and super-additive so that resource allocation is possible and provides SUs with an incentive to cooperate and form the grand coalition. The characteristic function of the game is derived based on the worths of SUs, calculated according to the amount of work done for the coalition in terms of reduction in uncertainty about PU activity. According to her worth in the coalition, each SU gets a pay-off that is computed using various one-point solutions such as Shapley value, $\tau$-value and Nucleolus. Depending upon their data rate requirements for transmission, SUs use the earned pay-off to bid for idle channels through a socially optimal Vickrey-Clarke-Groves (VCG) auction mechanism. Simulation results show that, in comparison with other resource allocation models, the proposed cooperative game-theoretic model provides the best balance between fairness, cooperation and performance in terms of data rates achieved by each SU.

\end{abstract}


\IEEEpeerreviewmaketitle

\section{Introduction}

Current wireless networks are characterized by static spectrum allocation policy, where spectrum is assigned to license holders on a long term basis. Due to recent increase in spectrum demand, certain bands face severe scarcity and yet, a large portion of the spectrum is often under-utilized across time and space \cite{fcc}. The apparent scarcity in spectrum arises from rigid frequency allocations rather than actual physical shortage of the spectrum. Techniques facilitating flexible spectrum usage have been developed in order to solve these inefficiency problems. The key enabling technology of dynamic spectrum access techniques is cognitive radio (CR) technology \cite{167416}, \cite{1391031}, \cite{1039509}, which provides the capability for unlicensed secondary users (SUs) to opportunistically access unused licensed bands (spectrum overlay approach) without causing harmful interference to primary users (PUs).

In a CR network, SUs collaboratively sense the spectrum \cite{5072368}, \cite{1447503}, based on a sensing policy to identify available portions of the spectrum referred to as \textit{spectrum holes}. By exchanging information about the state of spectrum occupancy in terms of local log likelihood ratios (LLRs) and side information such as signal to noise ratios (SNRs), SUs are able to improve detection performance and network coverage \cite{554208}. After sensing, SUs share the available spectrum amongst themselves and coordinate transmission attempts in idle channels based on a access policy. A Fusion Center (FC) manages the coalition's sensing and access policy. Since spectrum access is entirely dependent on sensing results in a CR network, both policies are closely interlinked and hence, a joint spectrum sensing and access policy must be modeled to optimize the utilization of spectral resources. Thus, there is a need to normatively model the spectrum sensing and sharing problem jointly in CRs to design, regulate and evaluate the system performance.

Sharing the benefits of cooperative sensing is a non-trivial problem and is of great interest. In order to arrive at an access policy that is acceptable to all SUs, the competition and conflict of interest between the SUs must be resolved in a fair manner. This calls for a game-theoretic approach to modeling the problem at hand. For example, if the identification of unoccupied spectrum bands by the SUs can be construed as \textit{value} in a quantitative and/or qualitative sense, the problem of accessing unoccupied spectrum bands reduces to allocating this \textit{value} among the SUs by means of a cooperative game. A cooperative game \cite{osbrub}, \cite{Bradimtij}, focuses on what groups of SUs (coalitions) can jointly achieve without considering how the SUs function internally in the coalition. The main assumption in cooperative games is that the grand coalition of all SUs within a certain geographical area will form and hence, the aim of the game is to allocate the overall value created by the SUs. Thus, modeling the spectrum sensing and sharing problem in CRs as a cooperative game is both intuitively and logically appealing. 

It is possible that the spectral resources are quite limited in order to be allocated to all SUs or the SUs themselves might not need to access the channel immediately but at a later time after sensing. Hence, it would be useful to have a transferable utility (TU) where the SUs can translate the allocations to some form of common currency and depending upon their data rate requirements for transmission, use the currency to bid on idle channels. A suitable mechanism is needed to facilitate the process of coordinating the bids from SUs and allocating idle channels to them. Vickrey-Clarke-Groves (VCG) auction \cite{vickrey-61}, \cite{Clarke-Groves}, based approach is considered here in order to demonstrate that a socially optimal and feasible mechanism exists that can allocate idle channels to SUs. However, devising an optimal bidding strategy for SUs is a complex combinatorial problem that entails a separate study in mechanism design and is beyond the scope of this paper. 

Non-cooperative game theory has been used to study, design, model and evaluate the performance of CRs in \cite{5230850}, \cite{5426522}, \cite{4570231}. Bargaining solutions to spectrum sharing problem in CRs have also been studied in the past \cite{4907475}. A detailed tutorial on coalitional game theory for communication networks can be found in \cite{saad}. Cooperative game theory has been applied in higher layers such as network and transportation layers to study routing protocols \cite{Cai2004}, \cite{4289458}, to study packet forwarding \cite{4768588}, and to study resource allocation \cite{1457343}. Coalitional game theory has been used to model cooperation in wireless networks in \cite{5285181}. Auctioning has been studied in detail with respect to spectrum sharing in \cite{spectrumauction}. The authors are not aware of any joint modeling of spectrum sensing and sharing in CRs using game theory or other methods.

In this paper, we propose a framework for joint modeling of the spectrum sensing and sharing problem in CRs as a cooperative game with transferable utility. The worth of individual players/coalitions in the game is calculated according to the quality and quantity of work done for the coalition. The resulting games have an inherent structure that provides the games with desirable properties such as balancedness and super-additivity. Balanced games have non-empty cores which makes resource allocation possible and stable. Super-additive games ensure that grand coalition is always formed and that the SUs have an incentive to cooperate rather than compete with each other. Since the core could be very large, one-point solutions such as Shapley value, $\tau$-value and Nucleolus are computed that provide singleton pay-off allocations. Depending upon their data rate requirements, SUs bid accordingly on idle channels with pay-offs obtained from cooperatively sensing the spectrum. A Vickrey-Clarke-Groves (VCG) auction is used to allocate resources to SUs to demonstrate that a feasible mechanism exists that can allocate idle channels to SUs based on their bids and data rate requirements.

The contributions of this paper are as follows :

\begin{itemize}

\item A novel and comprehensive framework for normative joint modeling of the spectrum sensing and sharing problem in CRs as a transferable utility (TU) cooperative game is proposed.

\item The characteristic function of the game is derived based on the worths of each SU calculated according to the amount of work done by the SU for the coalition.

\item The quantity and quality of work done by the SU is measured in terms of reduction in uncertainty about the PU that the SU brings from sensing the channels.

\item The resulting game is shown to be balanced and super-additive in nature thereby ensuring that the resource allocation is possible and that SUs have an incentive to cooperate.

\item Various one-point solutions such as the Shapley value, $\tau$-value and Nucleolus are computed that provide singleton pay-offs to the SUs.

\item A VCG auction mechanism is considered as an example to demonstrate that a socially optimal feasible mechanism exists that can allocate idle channels to SUs based on their bids and data rate constraints for transmission.

\item Examples that illustrate the \textit{proof of concept} of the proposed game-theoretical allocation mechanism are provided.

\item The cooperative game-theoretic method is compared against other common allocation models to show that the proposed method achieves the best balance between fairness, cooperation and performance in terms of data rate achieved by each SU.

\end{itemize}

The rest of this paper is organized as follows. The CR network structure and the model used for spectrum sensing and sharing in CRs is over viewed in Section II. The proposed game-theoretic modeling and VCG auctioning is described in detail in Section IV. In particular, spectrum sensing and sharing in CRs is normatively modeled as a cooperative game, the characteristic function of the game is derived, pay-offs to SUs are computed, the game is characterized and the mechanism of VCG auction is explained. Simulation examples that show how the proposed model allocates resources in a fair and stable manner are illustrated in Section V. The proposed cooperative game-theoretic model is also compared against other common allocation models and the performance in terms of data rate achieved by each SU is analyzed. Section VI concludes the paper. Analytical results that prove the various properties of characteristic function of the modeled game such as non-negativity, monotonicity, balancedness and super-additivity are provided in the Appendix.

\section{System Model}

In this section, we describe the CR network structure and the model used for spectrum sensing and sharing in CRs.

CR networks enable efficient spectrum usage for SUs via dynamic spectrum access techniques and heterogeneous networks without causing harmful interference to PUs. Multiple SUs equipped with CR enabled devices form centralized (CR base station) or ad-hoc network structures to perform spectrum management functions such as spectrum sensing, spectrum decision, spectrum sharing and interference management \cite{4481339}. 

The available spectrum may consist of contiguous or scattered frequency bands and is assumed to be subdivided into $M$ sub-bands. There are one or more PUs operating in these sub-bands. Without any loss in generality, PU activity can be assumed to be random as there is no a priori information. If there is a prior model for PU activity, it can be incorporated into the game without affecting the system model. There are $N$ SUs who wish to cooperate and make use of the unused licensed spectrum. A FC manages the sensing and access policy of the SUs. In the absence of a dedicated central entity, one of the SUs can serve as the FC. The FC and SUs communicate with each other through a common control channel. SUs sense the channels for PU activity based on any one of the available sensing methods such as energy detection, cyclostationary-based detection, matched filter etc. Decision statistics such as LLRs or other sufficient statistic combined with side information such as SNRs may be translated to probability of detecting ($P_d$) the PU in the channel. Alternatively, the SUs could send only the LLRs and the FC can translate them to probability of detecting the PU. Based on the probabilities of detection of PU activity reported by SUs, FC makes a decision about channel availability. The soft decision statistics reported by the SUs enable the FC to make a qualitatively better decision about spectrum occupancy.

Once the idle channels have been identified, the worth of each user and each coalition is calculated by the FC based on the quantity and quality of work done. For example, the quantity of work done can be measured by the number of channels sensed and power spent in sensing them, while the quality of work can be measured in terms of reduction of uncertainty about PU activity. The worth of each user/coalition determines how the overall value created by the grand coalition is distributed among the SUs. Pay-offs received by each SU is calculated according to cooperative game-theoretic solution concepts. With this pay-off, SUs bid for idle channels according to their data transmission rate requirements or save the pay-offs for later use. FC acts as the auctioneer and allocates idle channels to bidders according to VCG auction mechanism.

\begin{figure}[htpb]
\centering
\includegraphics[width=0.48\textwidth]{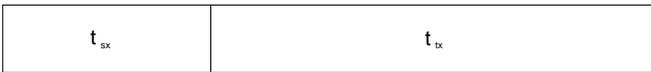}
\caption{Time slots are divided into sensing and transmission minislots.}
\label{timeslot}
\end{figure}

Every time slot $t$ consists of a sensing (Sx) minislot followed by a transmission (Tx) minislot as shown in Fig. \ref{timeslot}. 

Sensing periods are synchronized between SUs and simultaneous sensing and transmission in the same channel is not possible. In order to minimize the number of times each SU communicates with the FC during every time slot $t$, information exchange between them is modeled in the following manner. 

At the end of Sx minislot in $t$, each SU sends FC the following information :

\begin{itemize}

\item Sensing results in terms of LLRs or probability of detecting PU for the just concluded Sx minislot in $t$.

\item Estimated achievable capacities of the channels sensed in $t$ (based on instantaneous SINR).

\item A sealed bid for accessing idle channels during the upcoming Tx minislot in $t$ depending upon data rate requirements of the SU.

\item Preference for number of channels to be sensed by the SU during Sx minislot in $t+1$, after taking into consideration various factors such as type of data to be transmitted, data rate, QoS, power constraints, etc.

\end{itemize}

At the beginning of Tx minislot in $t$, FC sends SUs the following information :

\begin{itemize}

\item A Channel allocation map indicating which SU gets to transmit over which idle channel during Tx minislot in $t$ based upon the SU bids and estimated channel capacities.

\item Normalized pay-offs received by SUs for sensing the spectrum up until $t$ slots with which SUs bid at the end of Sx minislot in $t+1$ for idle channels that will be available at time interval $t+1$.

\item A spectrum sensing allocation map according to which SUs sense different sub bands of spectrum during the Sx minislot in $t+1$.

\end{itemize}

It is also possible that the FC and the SUs could communicate with each other a few more times in slot $t$ to optimize the bidding procedure. Instead of bidding based on their own data rate requirements, the SUs could first obtain information about the idle channels and then make an informed bid based on the estimated capacities of the idle channels. For instance, if the estimated capacities of idle channels are not good enough for the SU's data rate requirements, then she can make a low bid and save her pay-off for future bidding when the channel conditions are good. In further complex systems, SUs can learn the bidding strategy of other SUs or model the data rate requirements of other SUs and make a bid after adjusting for the newly available information. In addition to complicating the communication between the SUs and the FC, these methods consume additional power, create a lot of overhead on the control channel and encroach upon the time that the SUs could otherwise have spent in accessing the idle channels. If the gain is sufficiently high, the additional power, bandwidth and delay can be justified, but in this paper, a basic communication model between the SUs and the FC as described above is considered to avoid complexities.

\section{Game Modeling}

In this section, we describe in detail how the characteristic function of the proposed game is modeled, or in other words, how the worth of each individual user and coalitions are calculated at the FC and how their respective pay-offs are derived. Additionally, we also describe the Vickrey-Clarke-Groves (VCG) auctioning procedure in a cooperative game-theoretic setting to demonstrate that a feasible mechanism exists to allocate unoccupied bands of the spectrum to cooperating SUs based on their pay-offs and data rate requirements.

Since there is no information or model about PU activity, it is assumed to be random and is modeled as a binary random variable. Thus, the average probability of detecting the PU before sensing the channel without any a priori information is 0.5. The binary entropy function $H(P_d)$ measures the amount of uncertainty associated with the detection probabilities $P_d$. In other words, the uncertainty associated with the PU activity before sensing the channel is at its maximum with a value of $H(0.5) = 1$. After sensing the channel, SNRs are translated to probability of detection of the PU occupying the channel. These detection probabilities, irrespective of the sensing method used, help to reduce the uncertainty about the PU activity after sensing the channel.

The key principle involved in calculating the worth of each SU is based on the reduction in uncertainty about the PU activity that she brings from sensing the channels. This reduction in uncertainty is calculated as $H(0.5) - H(p_{ij})$ = $1 - H(p_{ij})$, where $p_{ij}$ is the probability of detection of PU by SU $i$ on channel $j$. However, the SU is not rewarded if its detection probability value is not in agreement with the global decision taken by the FC about PU activity. Moreover, there might be other users and coalitions (entities in general) that are sensing the particular channel and bringing in information to the FC about PU activity on that channel. Since the FC values the information from all entities equally, the information brought in by a SU will be appropriately weighted by the total number of entities sensing this channel. Thus, the worth of the SU $i$ on channel $j$ is given by,
\begin{equation} v(\{i\}) = \frac{1-H(p_{ij})}{c_i(j)}, \label{eq1} \end{equation} 
where $c_i(j)$ is the total number of entities sensing channel $j$ in addition to SU $i$. Aggregating the reduction in uncertainty from all sensed channels, the worth of the SU is given by,
\begin{equation} v(\{i\}) = \sum\limits_{j=1}^{M} \frac{1-H(p_{ij})}{c_i(j)}, \label{eq2} \end{equation} 
where $M$ is the total number of channels.

Calculating the worth of a coalition is slightly more cumbersome. The coalition is treated as if it were a single user. The detection probability values for the coalition are chosen from SUs within the coalition in such a way that the best detection probability among the SUs is selected for the coalition. Since the coalition is bound by the FC's decision on spectrum occupancy, the best detection probability for the coalition is deemed to be the probability that agrees as closely as possible with the decision taken by the FC about PU activity, thereby maximizing the worth of the coalition. The benefits of cooperation are quite evident in this set-up where the worth of the coalition is maximized and all SUs of the coalition benefit by the best performing SU within the coalition who is proportionally rewarded for her performance. Thus, the best detection probability value for the coalition is given by $| \ \underset{\forall i \in S}{max}\left(p_{ij}.D_j\right)|$, where $D_j$ is the spectrum decision on channel $j$ taken by the FC (+1 when PU is present and -1 when PU is absent). Due to diversity gains, information obtained from a coalition is more reliable than the information obtained from a single SU. In order to account for this, the total reduction in uncertainty about the PU activity brought in by the coalition is appropriately weighted by number of SUs in the coalition.

The general characteristic function is thus given by,
\begin{equation}
v(S) = |S| \ \sum\limits_{j = 1}^{M} \left(\frac{1 - H\left(| \underset{\forall i \in S}{max}\left(p_{ij}.D_j\right)|\right)}{c_S(j)}\right) \ , \label{equation}
\end{equation}
where, $S$ is any coalition in $\{1,2,...,N\}$, $|S|$ represents the cardinality of the set $S$, $M$ is the number of channels, $H$ is the binary entropy function , $p_{ij}$ is the probability of detection of PU by SU $i$ on channel $j$, $D_j$ is the spectrum decision (+1 when PU is present and -1 when PU is absent) on channel $j$ and $c_S(j)$ is the total number of entities sensing channel $j$ including the coalition $S$.

Games modeled in this fashion are balanced and super-additive in nature. See Appendix for detailed proof. Balancedness ensures that the core of the game is non-empty and hence resource allocation (spectrum sharing) is possible and stable. Various one-point solutions can be calculated that provide unique pay-offs, especially the nucleolus, which is guaranteed to lie in the core as the core is non-empty \cite{pelsud}. Super-additivity ensures that no individual SU or a coalition of SUs deviates from the grand coalition and that the SUs prefer to form the grand coalition for cooperatively sensing and sharing the spectrum. However, from the FC's point of view, it might be beneficial if some SUs form a new coalition due to the prevailing channel conditions or location of PUs, but addressing these issues is out of scope of this paper. The absence of an incentive to split away from the grand coalition reassures the SUs a sense of fairness in the calculation of pay-offs.
 
The FC can proportionally allocate idle bands of the spectrum in the upcoming Tx minislot based upon the pay-offs earned by the SUs. However, it is possible that the spectrum resources are quite limited in order to be distributed to all SUs or the SUs themselves might not wish to transmit data in the upcoming Tx minislot. Hence, instead of directly translating the pay-offs to spectrum resources, it would be advantageous to have a system where the pay-offs can be transferred to some form of currency and be used according to the wishes of the SUs.

In a TU game, the pay-offs obtained from cooperatively sensing the spectrum are normalized and directly translated to a currency unit that can be used by the SUs to bid on idle channels. The SUs use these normalized pay-offs as currency to bid according to their data rate requirements for transmission. Thus, SUs with full buffers or an urgent need to transmit (for e.g., delay intolerant data such as video calls) will bid almost all of their pay-off to acquire the best possible channel for transmission, while SUs having a less urgent need to transmit (for e.g., delay tolerant data such as text messages) will judiciously bid only a part of their pay-off and reserve rest of the pay-off to bid when a greater demand to transmit arises in the future and SUs who do not wish to transmit need not make a bid at all. Therefore, we consider the VCG auctioning procedure to prove that a feasible mechanism exists that can allocate idle channels to SUs based on their pay-offs and data rate requirements.

The FC acting as the VCG auctioneer arranges bids from SUs and the highest bidding SU is allocated a channel that is best for her (in terms of estimated channel capacity) and is charged a price equal to the \textit{second highest bid plus a bid increment}. The bid increment ensures that the winner of the auction always pays a slightly higher price than the second highest bidder. This price is now subtracted from the bid of the SU who was just allocated the channel and bids from SUs are rearranged to allocate the remaining idle channels in a similar manner. This procedure continues until the FC runs out of idle channels to allocate. The balance pay-offs of SUs after the auction are normalized again and are averaged with the normalized pay-offs obtained from sensing the spectrum in the previous time slots. Normalized averaging of the pay-off ensures that no single SU can collect a huge amount of pay-off by not bidding for a long time and then suddenly hog the auction by exploiting all available spectrum opportunities and denying other SUs a fair chance of accessing the spectrum. This procedure of cooperatively sensing the spectrum, earning pay-offs through a transferable utility cooperative game, bidding on idle channels through a VCG auction is repeated over all time slots.

From the data rate point of view, by allocating to the highest bidder the best channel as seen by her at the price of the second highest bid, VCG auction maximizes total utilities of SUs and is hence referred to as a socially optimal mechanism. Also, VCG auction provides SUs with a weakly dominant strategy to truthfully bid according to the SU's actual transmission requirements. By allocating channels to SUs who value them most and have worked qualitatively and quantitatively to achieve a larger pay-off, the resulting game is construed to be fair by all SUs which provides them with an incentive for future cooperation.

\section{Simulation Examples}

In this section, we provide an example to illustrate the proposed cooperative game-theoretic approach to jointly modeling the spectrum sensing and accessing scenario in cognitive radios. A multi-band multi-user scenario is considered. We show that through the cooperative game-theoretic approach, a solution exists for the resource allocation problem and that the allocation is fair, stable, provides SUs with an incentive to cooperate and is socially optimal. We also compare the proposed model to other commonly known allocation models in terms of the rate achieved by each SU and the sum rate of all SUs to show that the proposed method provides the best balance between fairness, cooperation and performance.

Simulations were carried out in MATLAB with the help of TUGlab toolbox \cite{TUGlab}. The PU signal used in the simulations is the OFDM signal which is employed in many wireless communication systems such as 3GPP Long term evolution (LTE), IEEE 802.11 a/g/n  Wireless local area networks (WLAN), IEEE 802.16 Wireless metropolitan area networks (WMAN), Digital video broadcasting (DVB) standards DVB-T and DVB-T2. 

Let $H_0$ be the null hypotheses, i.e., an OFDM based PU is absent and $H_1$ be the alternate hypotheses, i.e., an OFDM based PU is present. The autocorrelation property of OFDM systems using cyclic prefix (CP) is used for the detection of PUs. Let the length of the useful symbol data $T_d = 32$ and the length of the cyclic prefix $T_{cp} = T_d/4 = 8$. The SUs employ a  autocorrelation based detector and the detection period is assumed to be 100 OFDM blocks. Therefore, the number of samples for the autocorrelation estimate at the SU detector is $100(T_d + T_{cp}) = 4000$ Samples. This corresponds to a sensing time of approximately 0.6 ms for a OFDM system with 7 MHz VHF TV channel bandwidth. Assuming that the test statistics $r$ are Gaussian distributed \cite{4804743} with zero mean and variance $\sigma^2$ under $H_0$ and Gaussian distributed with mean $\mu_1$ and variance $\sigma^2$ under $H_1$, for a fixed threshold $\eta$, the probability of false alarm is given by  $P_{fa} = P(r > \eta|H_0) = \frac{1}{2} \text{erfc} \left(\frac{\eta}{\sqrt{2}\sigma}\right)$, and the probability of detection is given by $
P_{d} = P(r > \eta|H_1) = \frac{1}{2} \text{erfc} \left(\frac{\eta-\mu_1}{\sqrt{2}\sigma}\right)$.

\begin{figure}[htpb]
\centering
\includegraphics[width=0.24\textwidth]{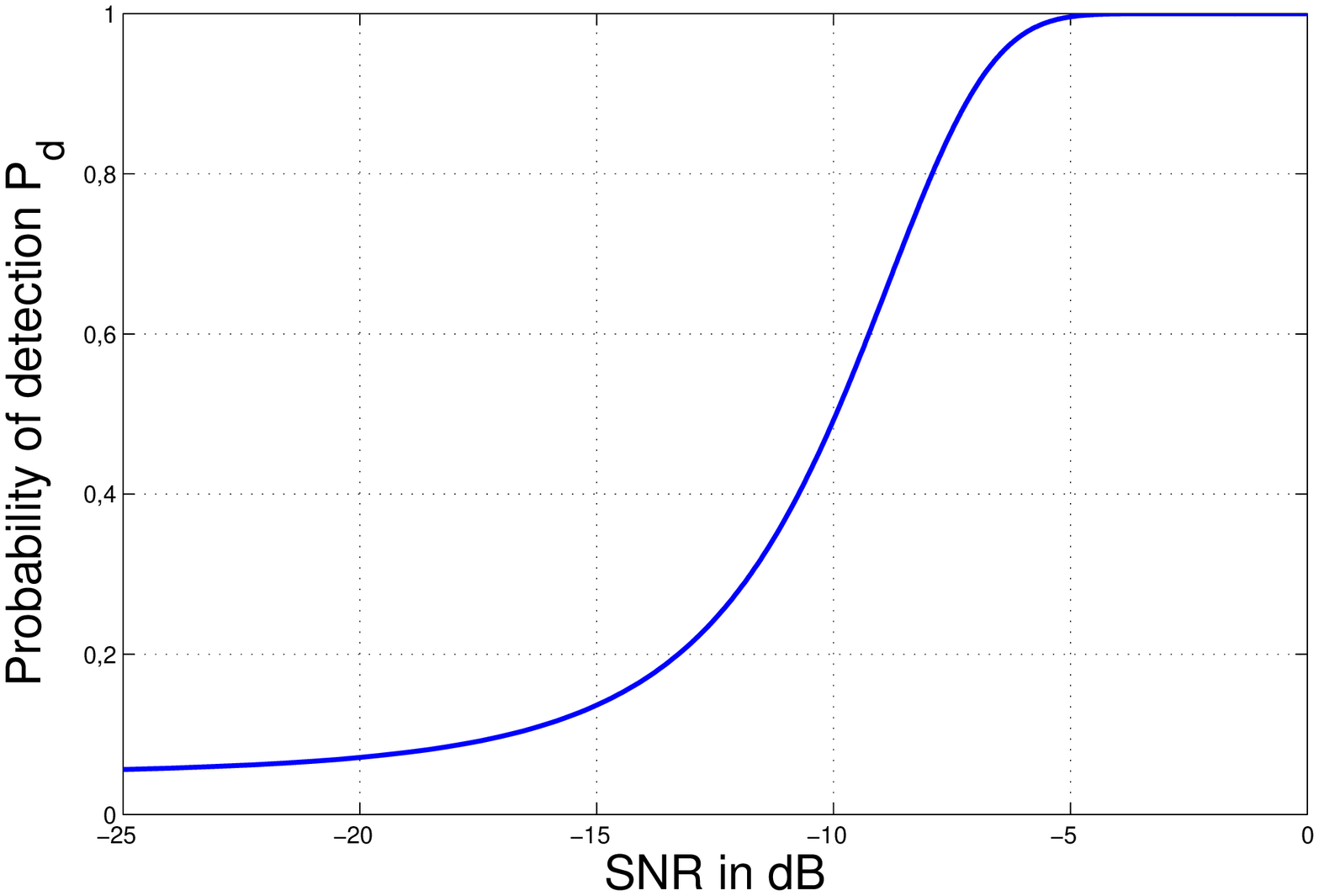}
\label{fig:a}
\hspace{-0.2in}
\includegraphics[width=0.24\textwidth]{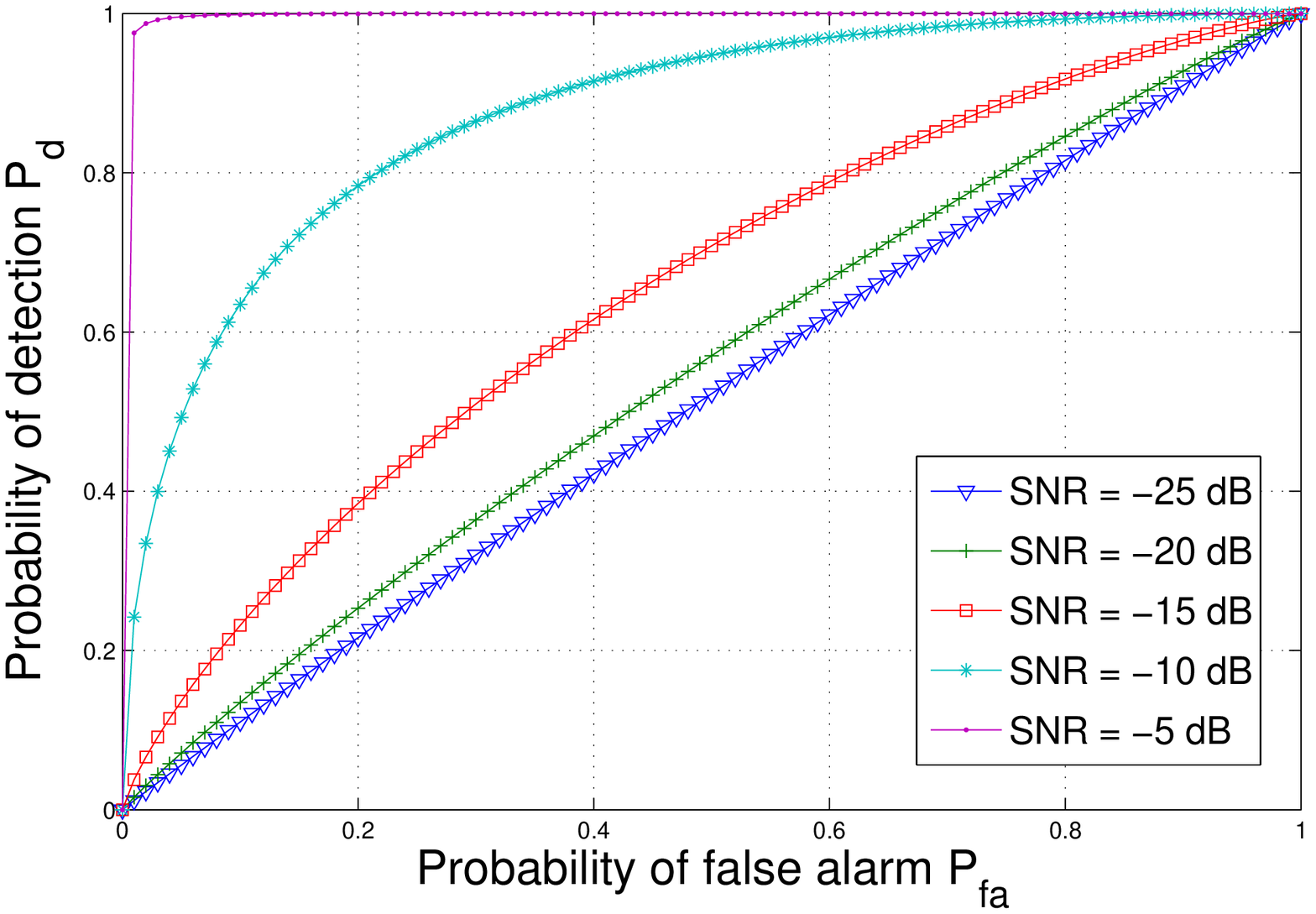}
\label{fig:b}
\caption{(a) SNR values are mapped to Probability of detection $P_d$ of PU for Neyman-Pearson detection under a constant false alarm rate $P_{fa} = 0.05$. (b) ROC plots for different values of SNRs.}
\label{fig:2}
\end{figure}

SUs employ a decentralized cooperative detection scheme where soft decision statistics from each SU is combined at the FC to detect PU activity based on a detection rule. In this case, assuming the parameters of OFDM based PU as mentioned earlier, the SUs sense the channel and translate SNRs to probability of detection $P_d$ of PU using Neyman-Pearson detection strategy under a fixed probability of false alarm $P_{fa} = 0.05$ as shown in Fig. \ref{fig:2}(a). The SNR is defined as SNR = $10 log_{10} \frac{\sigma^{2}_x}{\sigma^{2}_n}$, where $\sigma^{2}_x$ and $\sigma^{2}_n$ are the variances of the transmitted signal and noise respectively. The receiver operating characteristic (ROC) plots for the SU detector for various SNRs are shown in Fig. \ref{fig:2}(b).

\subsection{Proof of Concept}

In order to illustrate the proposed cooperative game theoretical approach using spectrum sensing and sharing in CRs as an example, let us consider the situation in which there are 3 SUs sensing 3 channels. SUs choose number of channels they prefer to sense according to their data rate requirements, power constraints etc. and the FC creates a spectrum sensing map accordingly as shown in Table \ref{tab:1}(a). The x in the table represent channels sensed by SUs. For channels that are sensed by SUs, SNR values are generated from a random variable that is uniformly distributed between -25 dB and -5 dB as shown in Table \ref{tab:1}(b).

\begin{table}[htpb]
\centering
\subtable[Sensing Map]{
\scriptsize{
\begin{tabular}{|cc|c|c|c|c|}
\cline{1-5}
\multicolumn{2}{|c|}{Sensing} & \multicolumn{3}{|c|} {\color{black}{Channels}} \\ \cline{3-5}
\multicolumn{2}{|c|}{Map} & \color{black}{1} & \color{black}{2} & \color{black}{3} \\ \cline{1-5}
\multicolumn{1}{|c|}{\multirow{3}{*}{\color{black}{SU}}} &
\multicolumn{1}{|c|}{\color{black}{1}} & \color{black}{x} & \color{black}{-} & \color{black}{-}     \\ \cline{2-5}
\multicolumn{1}{|c|}{}                        &
\multicolumn{1}{|c|}{\color{black}{2}} & \color{black}{-} & \color{black}{x} & \color{black}{x}    \\ \cline{2-5}
\multicolumn{1}{|c|}{}                        &
\multicolumn{1}{|c|}{\color{black}{3}} & \color{black}{x} & \color{black}{x} & \color{black}{-}    \\ \hline
\end{tabular}}
\label{tab:a}}
\hspace{-0.1in}
\subtable[SNR values]{
\scriptsize{
\begin{tabular}{|cc|c|c|c|c|}
\cline{1-5}
\multicolumn{2}{|c|}{SNR} & \multicolumn{3}{|c|} {\color{black}{Channels}} \\ \cline{3-5}
\multicolumn{2}{|c|}{Matrix} & \color{black}{1} & \color{black}{2} & \color{black}{3} \\ \cline{1-5}
\multicolumn{1}{|c|}{\multirow{3}{*}{\color{black}{SU}}} &
\multicolumn{1}{|c|}{\color{black}{1}} & \color{black}{-19.5949} & \color{black}{-} & \color{black}{-}     \\ \cline{2-5}
\multicolumn{1}{|c|}{}                        &
\multicolumn{1}{|c|}{\color{black}{2}} & \color{black}{-} & \color{black}{-7.2246} & \color{black}{-17.0642}    \\ \cline{2-5}
\multicolumn{1}{|c|}{}                        &
\multicolumn{1}{|c|}{\color{black}{3}} & \color{black}{-8.5656} & \color{black}{-17.1763} & \color{black}{-}    \\ \hline
\end{tabular}}
\label{tab:b}}
\caption{Spectrum sensing map created by fusion center and SNR values obtained by secondary users from sensing the channel}
\label{tab:1}
\end{table}
\vspace{-0.15in}

The obtained SNR values are translated to probability of detecting the PU and the SUs send the resulting soft detection probability values to the FC where they are combined and the idle channels are determined as shown in Table \ref{tab:2}. In this example, a simple OR-based decision rule was used. The presence of PU is denoted by +1, whereas -1 denotes the absence of PU.

\begin{table}[htpb]
\centering
\begin{tabular}{|cc|c|c|c|c|}
\cline{1-5}
\multicolumn{2}{|c|}{$P_d$} & \multicolumn{3}{|c|} {\color{black}{Channels}} \\ \cline{3-5}
\multicolumn{2}{|c|}{Matrix} & \color{black}{1} & \color{black}{2} & \color{black}{3} \\ \hline
\multicolumn{1}{|c|}{\multirow{3}{*}{\color{black}{SU}}} &
\multicolumn{1}{|c|}{\color{black}{1}} & \color{black}{0.0734} & \color{black}{0.5} & \color{black}{0.5}     \\ \cline{2-5}
\multicolumn{1}{|c|}{}                        &
\multicolumn{1}{|c|}{\color{black}{2}} & \color{black}{0.5} & \color{black}{0.8837} & \color{black}{0.0968}    \\ \cline{2-5}
\multicolumn{1}{|c|}{}                        &
\multicolumn{1}{|c|}{\color{black}{3}} & \color{black}{0.7054} & \color{black}{0.0953} & \color{black}{0.5}    \\ \hline
\multicolumn{2}{|c|}{\multirow{1}{*}{Decision}} &
\multicolumn{1}{|c|}{-1} & 1 & -1     \\ \hline
\end{tabular}
\caption{Determination of idle channels from detection probability values by OR-based decision rule.}
\label{tab:2}
\end{table}
\vspace{-0.15in}

\begin{figure}[htpb]
\centering
\includegraphics[width=0.24\textwidth]{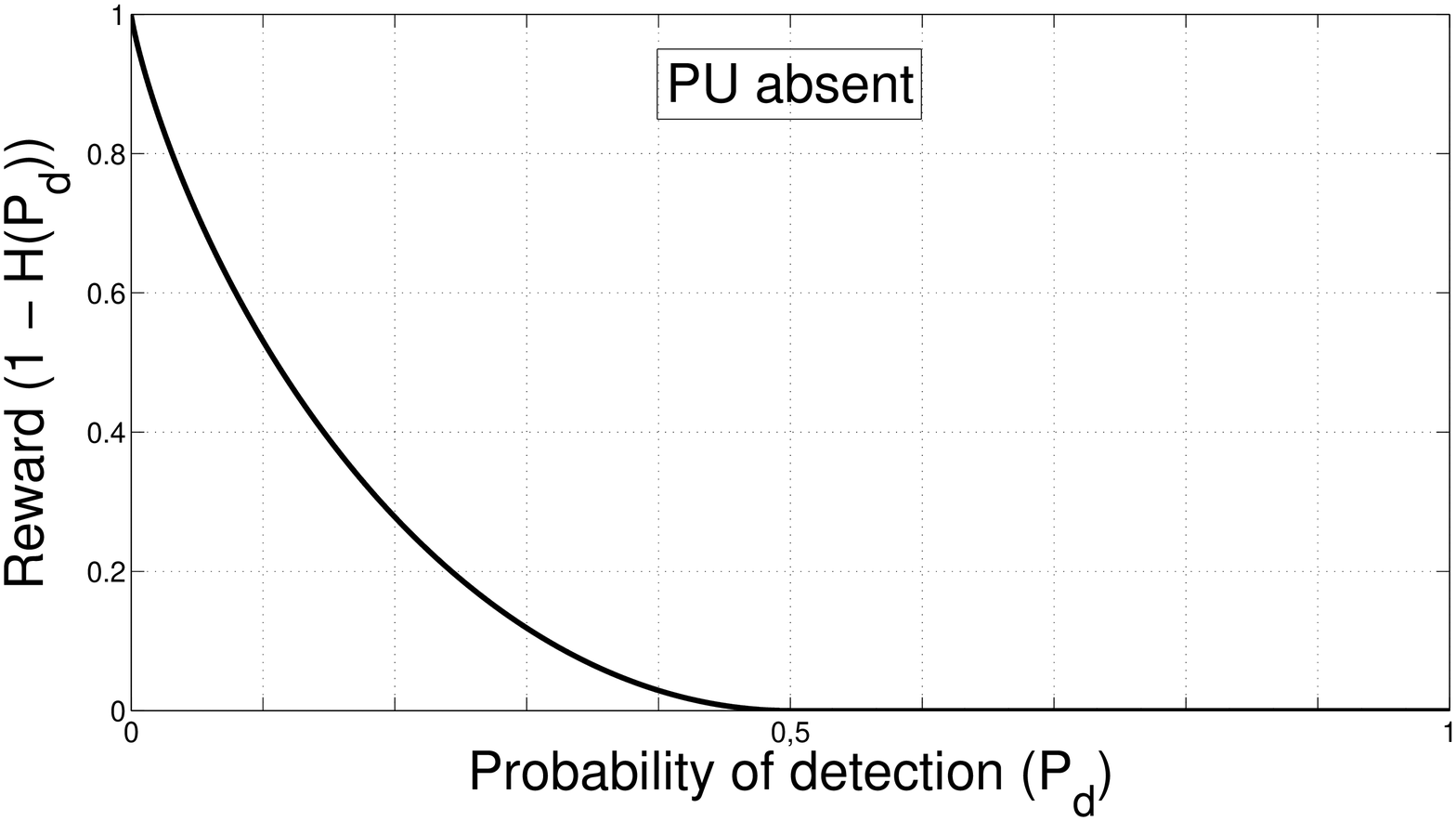}
\hspace{-0.2in}
\includegraphics[width=0.24\textwidth]{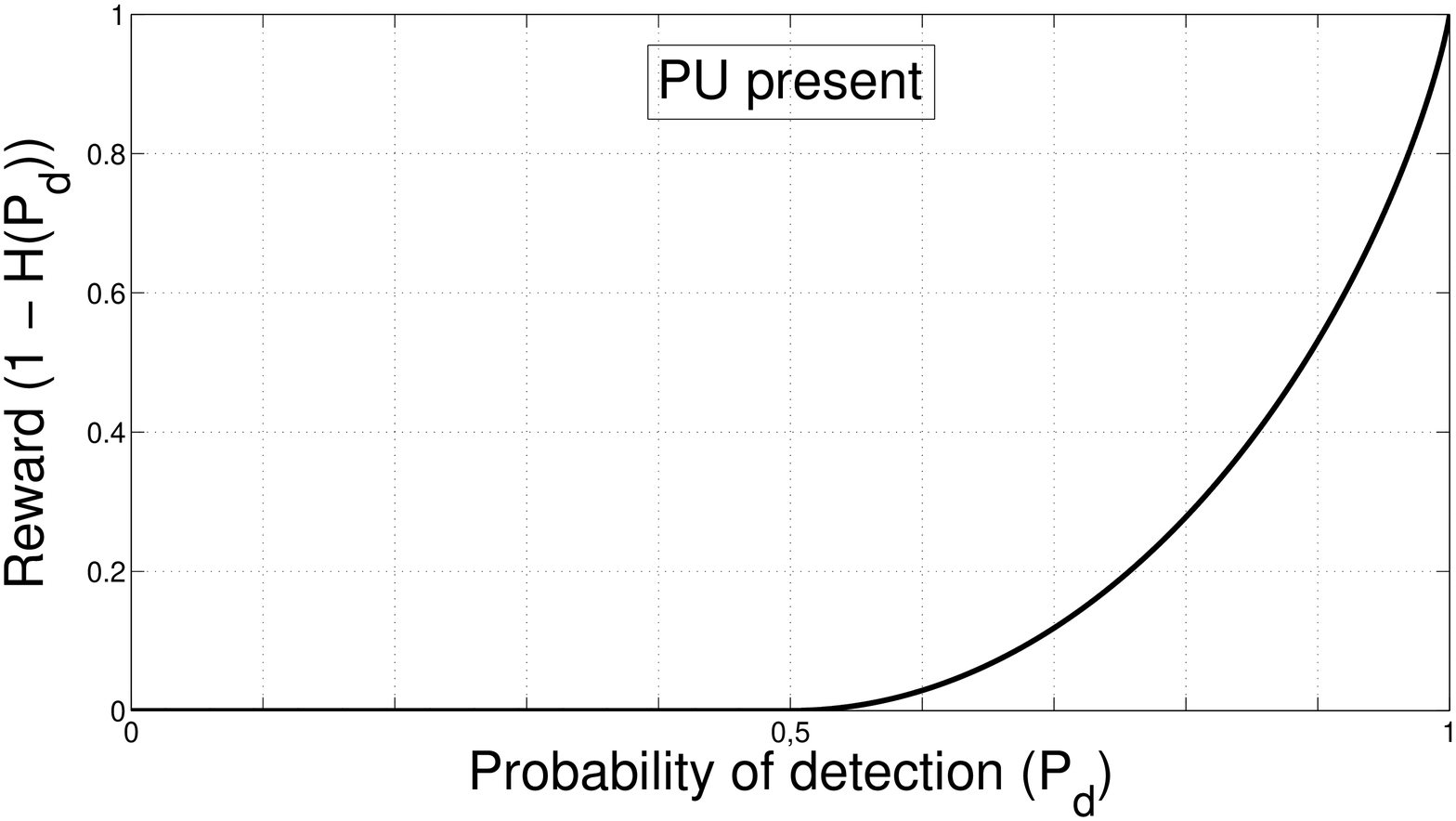}
\caption{Reward metric calculated in terms of reduction of uncertainty about PU activity from detection probability values after decision on PU activity is made. Note that SUs reporting detection probability values that contradict the decision taken by FC are not rewarded.}
\label{mivspd}
\end{figure}

Reduction in uncertainty about PU activity is used as the metric to reward SUs for the information that she brings from sensing the channel. However, it must be noted that this reward is conditioned on the decision about channel availability made by the FC as shown in Fig. \ref{mivspd}.

Using \eqref{equation}, the characteristic function for this game is calculated to be 
\begin{equation*}
\begin{split}
v &= \{v(1), v(2), v(3), v(12), v(13), v(23), v(123)\} \\ 
   &= \{0.3107, 0.7819, 0, 2.1851, 1.2427, 2.0450, 4.9316\} 
\end{split}
\end{equation*}

The characteristic function gives the worth of each SU and the worth of a SU in all possible coalitions. In this game, it can be seen that SU \#2 has the highest worth $v(2)$ for two reasons. First, SU \#2 sensed channel \#3 that was not sensed by any other user thereby bringing valuable information from that channel. Secondly, she was responsible for detecting the PU activity on channel \#2. Hence it is justified on the part of SU \#2 to expect the highest pay-off. Even though she has sensed only one channel, SU \#1 has the next highest individual worth $v(1)$ as her sensing result was in direct agreement with the decision taken at the FC about PU activity and expects to get the second highest pay-off. Though SU \#3 senses two channels, her sensing results were not in agreement with the decision taken at the FC about PU activity and hence ends up with an individual worth of 0. However, in coalition \{23\}, SU \#3 is the only user sensing channel \#1 and in coalition \{13\}, SU \#3 is the only user sensing channel \#2 thereby bringing valuable information from these channels. Since these contributions must be valued, it is fair on the part of SU \#3 to expect a non-zero pay-off even though her individual worth in the game is 0. From the characteristic function, three different one-point solutions for this game based on Shapley values, Tau values and Nucleolus are computed and shown in Table \ref{tab:3}.

\begin{table}[htpb]
\centering
\begin{tabular}{|cc|c|c|c|c|}
\cline{1-5}
\multicolumn{2}{|c|}{Normalized} & \multicolumn{3}{|c|}{One-point solutions} \\ \cline{3-5}
\multicolumn{2}{|c|}{pay-offs} & \color{black}{Shapley values} & \color{black}{Tau Values} & \color{black}{Nucleolus} \\ \cline{1-5}
\multicolumn{1}{|c|}{\multirow{3}{*}{\color{black}{SU}}} &
\multicolumn{1}{|c|}{\color{black}{1}} & \color{black}{30.5526} & \color{black}{30.6662} & \color{black}{32.2484} \\ \cline{2-5}
\multicolumn{1}{|c|}{}                        &
\multicolumn{1}{|c|}{\color{black}{2}} & \color{black}{43.4645} & \color{black}{43.3531} & \color{black}{41.8029} \\ \cline{2-5}
\multicolumn{1}{|c|}{}                        &
\multicolumn{1}{|c|}{\color{black}{3}} & \color{black}{25.9830} & \color{black}{25.9807} & \color{black}{25.9487} \\ \hline
\end{tabular}
\caption{Normalized pay-offs for the modeled cooperative game based on various one-point solutions.}
\label{tab:3}
\end{table}
\vspace{-0.15in}

From the normalized pay-offs shown in Table \ref{tab:3}, we see that the pay-offs reflect the expectations of the SUs. As stated earlier, games modeled in this fashion are balanced and super-additive in nature implying that a solution (pay-off) exists and that the SUs have an incentive to cooperate and form the grand coalition. The differences in the normalized pay-offs between the three one-point solutions arise by virtue of definition of these solutions. As a compromise, it is possible for the SUs to arrive at a single solution through bargaining, but the minor difference in pay-offs combined with the computational complexity renders the bargaining approach unnecessary. Moreover, it is guaranteed that the Nucleolus always lies in the core of the game ensuring the stability of the allocation and hence can be used as the default solution.

Working with normalized Nucleolus as pay-offs, SUs bid for channels based upon their transmission requirements which is modeled as a Gaussian random walk. The transmission requirements of the SUs can be quantified in terms of the size of the outgoing data buffer. Assuming additive white Gaussian noise (AWGN) channel with SNR values drawn from a uniform distribution between -25 dB and -5 dB, channel capacity (in Mbps) for a 7 MHz bandwidth is estimated using Shannon's capacity formula \cite{9171953} as shown in Table \ref{tab:4}.

\begin{table}[htpb]
\centering
\begin{tabular}{|cc|c|c|c|c|c|}
\cline{1-6}
\multicolumn{2}{|c|}{Channel} & \multicolumn{3}{|c|} {\color{black}{Channels}} & \multicolumn{1}{|c|}{SU} \\ \cline{3-5}
\multicolumn{2}{|c|}{Capacity} & \color{black}{1} & \color{black}{2} & \color{black}{3} & \multicolumn{1}{|c|}{Bids} \\ \cline{1-6}
\multicolumn{1}{|c|}{\multirow{3}{*}{\color{black}{SU}}} &
\multicolumn{1}{|c|}{\color{black}{1}} & \color{black}{0.0547} & \color{black}{0.0429} & \color{black}{0.0974} & \color{black}{24.7943}  \\ \cline{2-6}
\multicolumn{1}{|c|}{}                        &
\multicolumn{1}{|c|}{\color{black}{2}} & \color{black}{0.7187} & \color{black}{0.0143} & \color{black}{0.4765} & \color{black}{6.9917}  \\ \cline{2-6}
\multicolumn{1}{|c|}{}                        &
\multicolumn{1}{|c|}{\color{black}{3}} & \color{black}{2.0485} & \color{black}{0.9998} & \color{black}{0.0318} & \color{black}{22.3673}   \\ \hline
\end{tabular}
\caption{Estimated channel capacities along with secondary user bids}
\label{tab:4}
\end{table}
\vspace{-0.15in}

The VCG auction takes place as follows. From Table \ref{tab:4}, SU \#1 has the highest bid and the best channel for her in terms of achievable data rate is channel \#3. She gets access to channel \#3 at the price of the second highest bidder (SU \#3) plus one bid increment totaling 22.3674. This price is deducted from the bid of SU \#1 and the new bid of SU \#1 is 2.4269. The bids are sorted again and the auction is continued. SU \#3 is the highest bidder now and the best channel for her is channel \#1. As SU \#2 is the second highest bidder now, SU \#3 pays a price of 6.9918 for channel \#3. After the completion of bidding and allocation, remaining pay-offs of the SUs are normalized and averaged with the normalized pay-offs obtained from the next time slot. The allocation procedure is depicted in Table \ref{tab:5}.

\begin{table}[htpb]
\centering
\begin{tabular}{|c|c|c|c|}
\cline{1-4}
\multicolumn{1}{|c|}{Allocation} & \multicolumn{3}{|c|} {\color{black}{Secondary Users}} \\ \cline{2-4}
\multicolumn{1}{|c|}{table} & \color{black}{1} & \color{black}{2} & \color{black}{3} \\ \cline{1-4}
\multicolumn{1}{|c|}{\color{black}{Norm. Pay-off}} & \color{black}{32.2484} & \color{black}{41.8029} & \color{black}{25.9487}     \\ \cline{1-4}
\multicolumn{1}{|c|}{\color{black}{Bids}} & \color{black}{24.7943} & \color{black}{6.9917} & \color{black}{22.3673}    \\ \cline{1-4}
\multicolumn{1}{|c|}{\color{black}{Price Paid}} & \color{black}{22.3674} & \color{black}{-} & \color{black}{-}    \\ \cline{1-4}
\multicolumn{1}{|c|}{\color{black}{Channel Allocated}} & \color{black}{\#3} & \color{black}{-} & \color{black}{-}    \\ \cline{1-4}
\multicolumn{1}{|c|}{\color{black}{Rate Achieved (in Mbps)}} & \color{black}{0.0974} & \color{black}{-} & \color{black}{-}    \\ \cline{1-4}
\multicolumn{1}{|c|}{\color{black}{Bids}} & \color{black}{2.4269} & \color{black}{6.9917} & \color{black}{22.3673}    \\ \cline{1-4}
\multicolumn{1}{|c|}{\color{black}{Price Paid}} & \color{black}{-} & \color{black}{-} & \color{black}{6.9918}    \\ \cline{1-4}
\multicolumn{1}{|c|}{\color{black}{Channel Allocated}} & \color{black}{-} & \color{black}{-} & \color{black}{\#1}    \\ \cline{1-4}
\multicolumn{1}{|c|}{\color{black}{Rate Achieved (in Mbps)}} & \color{black}{-} & \color{black}{-} & \color{black}{2.0485}    \\ \cline{1-4}
\multicolumn{1}{|c|}{\color{black}{Balance Pay-off}} & \color{black}{9.8810} & \color{black}{41.8029} & \color{black}{18.9569}    \\ \cline{1-4}
\multicolumn{1}{|c|}{\color{black}{Norm. Balance Pay-off}} & \color{black}{13.9876} & \color{black}{59.1767} & \color{black}{26.8357}    \\ \hline
\end{tabular}
\caption{Allocation table showing user pay-offs and bids and subsequent channel allocations based on VCG auction}
\label{tab:5}
\end{table}
\vspace{-0.15in}

A few interesting things can be noted here. Though SU \#3 had the lowest pay-off, her need for transmission was greater than SU \#2 as was reflected in her bid and hence, she still gets access to a channel. It must be noted that the price paid by a SU for accessing a channel is not based on the estimated channel capacity, but is totally dependent on the data rate requirements of other users to access the channel. This is why SU \#1 pays a far higher price for a channel with lower estimated capacity compared to SU \#3. SU \#2 bids conservatively and reserves her pay-off for future use. At the end of the allocation, we see that SUs \#1 and \#3 have a very low normalized balance pay-off compared to SU \#2 and will have to work hard in terms of qualitatively and quantitatively sensing the spectrum to be able to bid competitively in the future. VCG auction does not guarantee data rate maximization, but allocates resources according to SU bids and hence satisfies the data rate requirements of the SUs in a fair manner. Thus, we see that the proposed cooperative game-theoretic model for jointly modeling the spectrum sensing and sharing scenario in cognitive radios results in an allocation that is fair, socially optimal and provides SUs with an incentive to cooperate in the future.

In the following sections, we compare the performance of the proposed cooperative game-theoretic model for resource allocation with other common allocation models in terms of the data rates achieved by each SU and the sum data rate of all SUs. We show that the proposed model provides the best balance between fairness, cooperation and performance among all the other models considered. Assuming a similar set-up as described in this section, we carry out simulations over 1000 time slots with 3 SUs sensing 5 channels.

\subsection{Comparison with joint sensing and probabilistic access model}

\begin{figure}[htpb]
\centering
\includegraphics[width=0.48\textwidth]{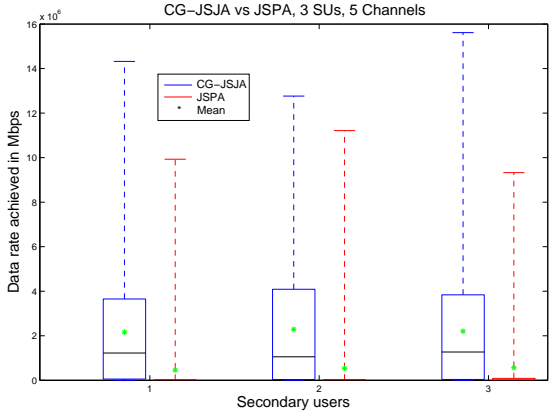}
\caption{Comparison between CG-JSJA and JSPA models. The box plot depicts the minimum value, first quartile, median, mean, third quartile and the maximum value of the achieved data rates for both models for each SU.}
\label{fig:3}
\end{figure}

The proposed cooperative game-theoretic joint sensing and joint access (CG-JSJA) model is compared with a joint sensing and probabilistic access model (JSPA). In JSPA, the SUs jointly sense the channel and collaboratively detect the PU at the FC, but access the idle channels in a probabilistic fashion very similar to carrier sense multiple access (CSMA) protocol. In order to compare the two models in a fair manner, in JSPA, the SUs choose the number of channels to access from the available free channels according to their data rate requirements for transmission which is proportional to their bids in the CG-JSJA model. From the channel capacity estimates that they have, SUs choose the channels with maximum capacity for transmission. During the transmit minislot, SUs listen to the channel and if found idle, transmit their data. In the event of a collision, the SUs back off for a random amount of time before retransmitting their data. The data rates achieved by each SU in both the models are shown as a box plot in Fig. \ref{fig:3}. It can be seen that each SU achieves a higher maximum, mean, median, lower quartile and upper quartile data rate in the cooperative game-theoretic approach when compared to the probabilistic access model. The lower performance of the JSPA model can be attributed to the collisions that occur among the SUs due to the uncoordinated nature of their transmissions. This shows that there is need for a joint access mechanism for the SUs to effectively making use of the idle channels and increase their individual data rate.

\subsection{Comparison with independent sensing and probabilistic access model}

\begin{figure}[htpb]
\centering
\includegraphics[width=0.48\textwidth]{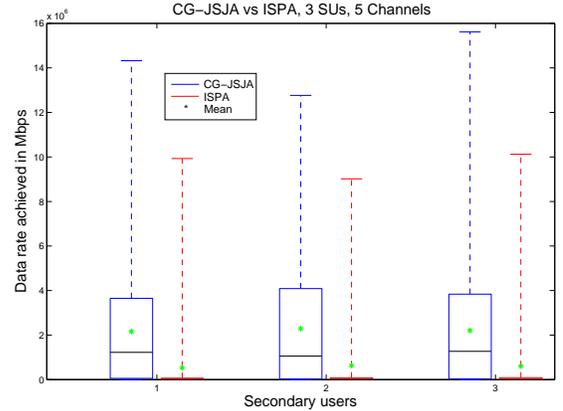}
\caption{Comparison between CG-JSJA and ISPA models. The box plot depicts the minimum value, first quartile, median, mean, third quartile and the maximum value of the achieved data rates for both models for each SU.}
\label{fig:4}
\end{figure}

The proposed CG-JSJA model is compared with a individual sensing and probabilistic access model (ISPA). In ISPA, the SUs sense the channels individually and detect the PU on their own and access  the idle channels in a probabilistic fashion very similar to CSMA protocol. As in the JSPA model, in ISPA too, the SUs choose the number of channels to access from the available free channels according to their data rate requirements which is proportional to their bids in the CG-JSJA model. During the transmit minislot, SUs listen to the channel and if found idle, transmit their data. In the event of a collision, the SUs back off for a random amount of time before retransmitting their data. The data rates achieved by each SU in both the models are shown as a box plot in Fig. \ref{fig:4}. Due to independent sensing on 5 channels, the 3 SUs failed to detect the PU activity 2605 times in 1000 time slots which could have been avoided by jointly sensing the channels. The lower performance of the ISPA model is due to the collisions occurring between the SUs during uncoordinated channel access, thereby demonstrating the necessity for both jointly sensing and accessing the channel cooperatively.

\subsection{Comparison with joint sensing and round robin model}

\begin{figure}[htpb]
\centering
\includegraphics[width=0.48\textwidth]{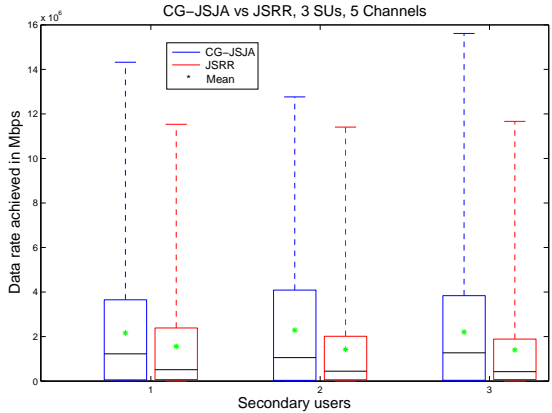}
\caption{Comparison between CG-JSJA and JSRR models. The box plot depicts the minimum value, first quartile, median, mean, third quartile and the maximum value of the achieved data rates for both models for each SU.}
\label{fig:5}
\end{figure}

The proposed CG-JSJA model is compared with a joint sensing and round robin access model (JSRR). In JSRR, the SUs jointly sense the channel and collaboratively detect the PU at the FC, but the FC allocates the idle channels in a round robin fashion. Since SUs take turns to transmit their data, all SUs get an equal opportunity in terms of accessing idle channels and there are no collisions between the SUs. However, in this model, the FC does not take into account the data rate requirements of the SU or suitability of the channel for the SU while making the allocation. The data rates achieved by each SU in both the models are shown as a box plot in Fig. \ref{fig:5}. From the SU's perspective, the JSRR model is fair enough to provide the SUs with an incentive to cooperate, but the socially optimal allocation in CG-JSJA model consistently achieves higher data rates over the JSRR model for all SUs.

\subsection{Comparison with joint sensing and rate maximization model}

\begin{figure}[htpb]
\centering
\includegraphics[width=0.48\textwidth]{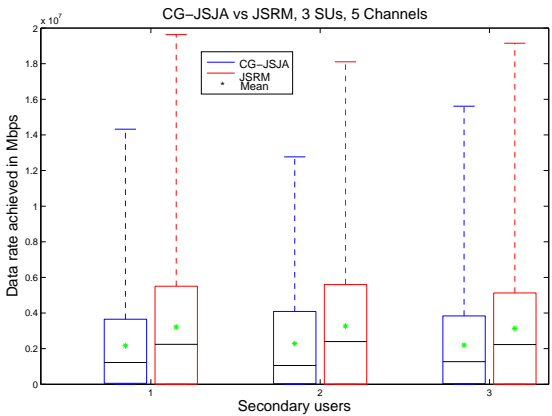}
\caption{Comparison between CG-JSJA and JSRM models. The box plot depicts the minimum value, first quartile, median, mean, third quartile and the maximum value of the achieved data rates for both models for each SU.}
\label{fig:6}
\end{figure}

The proposed CG-JSJA model is compared with a joint sensing and rate maximization access model (JSRM). In JSRM, the SUs jointly sense the channel and collaboratively detect the PU at the FC, but the FC allocates the idle channels in a manner that maximizes the overall data rate achieved. Therefore, an idle channel is allocated to the SU whose capacity estimate on that channel is the highest. Thus, the sum data rate is maximized and there are no collisions between the SUs, but the FC does not take into account the data rate requirements of the SUs. The data rates achieved by each SU in both the models are shown as a box plot in Fig. \ref{fig:6}. As expected, the JSRM model consistently outperforms the proposed CG-JSJA model.

\subsection{Comparison among all allocation models}

\begin{figure}[htpb]
\includegraphics[width=0.48\textwidth]{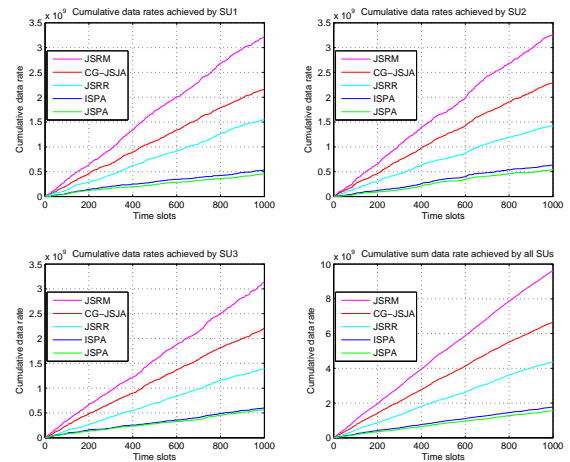}
\caption{Comparison between CG-JSJA, JSRM, JSRR, JSPA and ISPA models. The cumulative data rates achieved by each SU and the cumulative sum data rate achieved by all SUs is shown for all models.}
\label{fig:7}
\end{figure}

The cumulative data rates achieved by each SU in all the models are shown in Fig. \ref{fig:7}. It is interesting to note that the ISPA model performs slightly better than the JSPA model. This is because, in the ISPA model, when the SUs fail to detect the PU by themselves, they have more channels to choose from when randomly accessing the channels. Since the number of idle channels as perceived by each SU increases due to the missed detections of the PU, there are fewer collisions amongst the SUs. However, the increased data rate is achieved at the cost of interefering with the PU (2605 missed detections of PU by 3 SUs sensing 5 channels over 1000 time slots) and if the penalty associated with such missed detections of PUs is extremely high, the marginal increase in data rate of the ISPA model is not justified. As intended, the JSRM model achieves the highest possible data rate. The JSRR model, while providing SUs with an incentive to cooperate, is not able to match the data rate achieved by the CG-JSJA model.

\begin{figure}[htpb]
\includegraphics[width=0.48\textwidth]{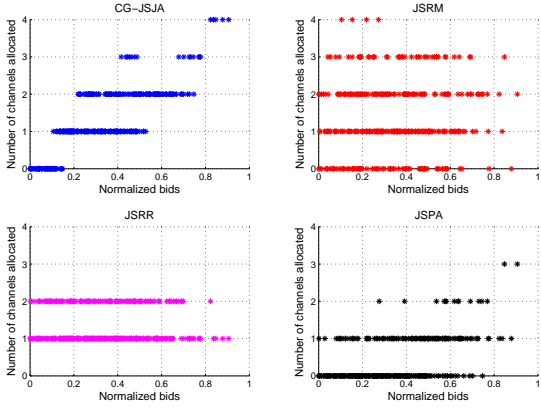}
\caption{Scatter plot depicting the number of channels allocated versus normalized bids. The plot shows the number of channels allocated to SU \#3 for every normalized bid made by SU \#3 in each time slot over 1000 time slots when the number of detected idle channels is equal to four for CG-JSJA, JSRM, JSRR and JSPA models. In the CG-JSJA model, the tiered step-like structure indicates that higher the bid made by the SU, the higher the probability of being allocated more channels. Since bids are made by SUs from pay-offs earned from sensing the channel, the amount of work done by SUs and their data rate requirements are fairly reflected in the number of channels allocated in CG-JSJA model. In comparison, other models are not construed by the SUs as fair due the fact that the number of channels allocated to them is independent of the bids made by them.}
\label{fig:8}
\end{figure}

The extent of fairness of these allocation models can be studied in terms of the number of channels accessed by each SU based on the bid that she makes. Fig. \ref{fig:8} shows the scatter plot for the number of channels allocated to SU \#3 for the normalized bid made by SU \#3 in each time slot over 1000 time slots when the number of detected idle channels is equal to four for all models. 

In the CG-JSJA model, it can be seen that the number of channels allocated is directly dependent on the bid made by the SU. The higher the bid made by the SU, the higher the probability of being allocated more channels. The SUs make their bids based on their data rate requirements for transmission and hence, the number of channels allocated to a SU is directly proportional to its data rate requirements for transmission. It must be noted here that the SUs cannot make high bids for the sake of being allocated more channels as the bids are based on the pay-offs they earn from sensing the channel. The only way for SUs to make high bids is through earning better pay-offs in terms of the quality and quantity of work done in sensing the channels. Thus, the CG-JSJA model is construed by the SUs to be a very fair model as both the data rate requirements of the SU and the amount of work done by the SU is reflected in the resource allocation without any room for manipulation by the SUs. 

For the JSRM model, it can be seen that the SU is allocated all the idle channels even with a very low bid while at another time slot, the SU is not allocated any channels even with a very high bid. The goal in JSRM model is to maximize the data rate and hence, the number of channels allocated to a SU is completely dependent on the channel conditions and independent of the bids made by the SU. Since the bids made by the SUs based on their data rate requirements for transmission are not taken into account while allocating channels, the JSRM model is not construed as a fair model by the SUs. The higher data rate achieved by the JSRM over the proposed CG-JSJA model is thus obtained at the cost of fairness. Similarly, in the JSRR model too, the data rate requirements of the SUs are not taken into account and it can be seen that even for very high bids, no SU gets access to more than $\left\lceil \text{number of idle channels/number of SUs} \right\rceil$ channels. However, the SUs are given an equal opportunity to access the channels in a round robin fashion which can be construed as somewhat fair despite the lack of consideration of the bids made by the SUs. In the JSPA or ISPA model, there is no allocation \textit{per se}, as the channels are accessed in a probabilistic fashion and it can be seen that no SU is able to access all the idle channels due to the collisions arising from uncoordinated channel access.

The comparisons between various allocation models are summarized in Table \ref{tab:6}.

\begin{table*}[htpb]
\centering
\small{
\begin{tabular}{|cc|c|c|c|c|c|}
\cline{1-7}
\multicolumn{2}{|c|}{Comparison} & \multicolumn{5}{|c|} {\color{black}{Model}} \\ \cline{3-7}
\multicolumn{2}{|c|}{Table} & \color{black}{CG-JSJA} & \color{black}{JSPA} & \color{black}{ISPA} & \color{black}{JSRR} & \color{black}{JSRM} \\ \cline{1-7}
\multicolumn{1}{|c|}{\multirow{3}{*}{\color{black}{Properties}}} &
\multicolumn{1}{|c|}{\color{black}{Fairness}} & \color{black}{Very High} & \color{black}{Low} & \color{black}{Low} & \color{black}{High} & \color{black}{Low} \\ \cline{2-7}
\multicolumn{1}{|c|}{}                        &
\multicolumn{1}{|c|}{\color{black}{Cooperation}} & \color{black}{Very High} & \color{black}{Moderate} & \color{black}{Low} & \color{black}{High} & \color{black}{High} \\ \cline{2-7}
\multicolumn{1}{|c|}{}                        &
\multicolumn{1}{|c|}{\color{black}{Achieved Rate}} & \color{black}{High} & \color{black}{Low} & \color{black}{Low} & \color{black}{Moderate} & \color{black}{Very High} \\ \hline
\end{tabular}}
\caption{Comparison between the proposed cooperative game-theoretic model and other allocation models. The proposed CG-JSJA model provides the best balance between fairness, cooperation and performance in terms of data rates achieved by each SU as well as the sum data rate achieved by all SUs.}
\label{tab:6}
\end{table*}

Thus, the proposed cooperative game-theoretic joint sensing and joint access model provides the best balance between fairness, cooperation and performance in terms of data rates achieved by each SU as well as the sum data rate achieved by all SUs.

\section{Conclusion}

In this paper, we have proposed a novel and comprehensive framework for normatively modeling spectrum sensing and sharing problem in cognitive radios as a transferable utility (TU) cooperative game. Additionally, Vickrey-Clarke-Groves (VCG) auction is used to demonstrate the existence of a feasible mechanism to allocate resources to secondary users (SUs). The characteristic function of the cooperative game is derived based on the worths of SUs, which is calculated according to the amount of work done for the coalition in terms of qualitatively and quantitatively obtaining information about the probability of detecting the primary user (PU). Games modeled in this fashion have desirable properties such as balancedness and super-additivity ensuring that the resource allocation is possible and providing SUs with an incentive to cooperate. Various one-point solutions such as Shapley value, $\tau$-value and Nucleolus are used to compute singleton pay-offs to the SUs, especially the Nucleolus, which lies within the core. SUs use these pay-offs to bid for idle channels based on their data rate constraints for transmission and estimated channel capacities. VCG auction provides SUs with a dominant strategy to truthfully bid their values, maximizes total utility of the SUs and allocates resources to bidders who value them the most, thereby resulting in a socially optimal allocation. In comparison to other allocation models, the proposed cooperative game-theoretic model provides a attractive trade-off among design goals in fairness, stability, cooperation and achievable data rates. Thus, spectrum resources are allocated in a fair and stable manner, ensuring that SUs have a very strong incentive in the future to cooperatively sense and share unoccupied bands of the spectrum without compromising on the data rates achieved by the SUs.

\appendix

The characteristic function of the proposed cooperative game $(N,v)$ is given by \eqref{equation}. In the following sections, we provide proofs for various desirable properties of the characteristic function such as Non-negativity, Monotonicity, Balancedness and Super-additivity.

\subsection{Non-negativity}

$\textbf{For each non-empty} \ \mathbf{S \in N, \ v(S) \geq 0.}$\vspace{0.1in}

Since $p_{ij}$ is the probability of detection of PU by SU $i$ on channel $j$, $0 \leq p_{ij} \leq 1$. Also, $d_j = \pm 1$. Therefore, $0 \leq | \underset{\forall i \in S}{max}\left(p_{ij}.d_j\right)| \leq 1$. Since $H$ is the binary entropy function, we have $0 \leq H\left(| \underset{\forall i \in S}{max}\left(p_{ij}.d_j\right)|\right) \leq 1$, which also implies that $0 \leq 1-H\left(| \underset{\forall i \in S}{max}\left(p_{ij}.d_j\right)|\right) \leq 1$. Also, $c_S(j) > 0$ and $|S| > 0$. Thus, we have,

\begin{equation}
|S|\sum\limits_{j=1}^{M}\left((1-H(|\underset{\forall i \in S}{max}\left(p_{ij}.d_j\right)|))/c_S(j)\right) \geq 0.
\label{2}
\end{equation}

Hence, $v(S) \geq 0$.

\subsection{Monotonicity}

$\textbf{For all non-empty} \ \mathbf{S, T \in N} \ \textbf{with} \ \mathbf{S \subset T, \ v(S) \leq v(T).}$\vspace{0.1in}

Let $d_j = +1$. Since the PU is present in the channel, only SUs with $p_{ij} \geq 0.5$ are rewarded. Therefore, $| \underset{\forall i \in S}{max}\left(p_{ij}.d_j\right)| \leq | \underset{\forall i \in T}{max}\left(p_{ij}.d_j\right)|$. For $p_{ij} \geq 0.5$, $H\left(| \underset{\forall i \in S}{max}\left(p_{ij}.d_j\right)|\right) \geq H\left(| \underset{\forall i \in T}{max}\left(p_{ij}.d_j\right)|\right)$. Let $d_j = -1$. Since the PU is absent in the channel, only SUs with $p_{ij} \leq 0.5$ are rewarded. Therefore, $| \underset{\forall i \in S}{max}\left(p_{ij}.d_j\right)| \geq | \underset{\forall i \in T}{max}\left(p_{ij}.d_j\right)|$. For $p_{ij} \leq 0.5$, $H\left(| \underset{\forall i \in S}{max}\left(p_{ij}.d_j\right)|\right) \geq H\left(| \underset{\forall i \in T}{max}\left(p_{ij}.d_j\right)|\right)$. Thus, in both cases, $1-H\left(| \underset{\forall i \in S}{max}\left(p_{ij}.d_j\right)|\right) \leq 1-H\left(| \underset{\forall i \in T}{max}\left(p_{ij}.d_j\right)|\right)$. Also, $c_S \geq c_T$ for all $S \subset T$. Since $|S| < |T|$, we have,

\begin{equation*}
|S|\sum\limits_{j=1}^{M}\left((1-H(|\underset{\forall i \in S}{max}\left(p_{ij}.d_j\right)|))/c_S(j)\right) \leq
\end{equation*}

\begin{equation}
|T|\sum\limits_{j=1}^{M}\left((1-H(|\underset{\forall i \in T}{max}\left(p_{ij}.d_j\right)|))/c_T(j)\right).
\label{3}
\end{equation}

Hence, $v(S) \leq v(T)$.

\subsection{Balancedness}

$\textbf{For each non-empty} \ \mathbf{S \in N, \ \sum_S \lambda_S v(S) \leq v(N)} \\ \textbf{for every balanced collection of weights} \ \mathbf{\lambda_S.}$\vspace{0.1in}

Let $R^S$ be the $S$ dimensional Euclidean space in which the dimensions are indexed by the members of $S$. The characteristic vector $1_S \in R^N$ of the coalition $S$ is given by

\begin{equation}
(1_S)_i = \left \{ \begin{array}{ll} 1 & \textrm{if~} i \in S \\ 0  & \textrm{otherwise.~} \end{array} \right. 
\label{4} 
\end{equation}

A collection $(\lambda_S)$ of numbers in $[0,1]$ is a balanced collection of weights, if for every player $i$, the sum of $\lambda_S$ over all the coalitions that contain $i$ is 1, such that $\sum_{S} \lambda_S 1_S = 1_N$. Since $S \subseteq N$, we have

\begin{equation*}
\sum\limits_{j=1}^{M} \left((1-H(|\underset{\forall i \in S}{max}\left(p_{ij}.d_j\right)|))/c_S(j)\right) \leq
\end{equation*}

\begin{equation}
\sum\limits_{j=1}^{M} \left((1-H(|\underset{\forall i \in N}{max}\left(p_{ij}.d_j\right)|))/c_N(j)\right). \ 
\label{5}
\end{equation}

Therefore, $\frac{1}{|S|}v(S) \leq \frac{1}{|N|}v(N)$, implying $v(S) \leq \frac{|S|}{|N|}v(N)$. Multiplying both sides by $\lambda_S$ and summing over all possible coalitions, we have, $\sum_S \lambda_S v(S) \leq \frac{v(N)}{|N|} \sum_S \lambda_S |S|$. Since the weights are balanced, $\frac{v(N)}{|N|} \sum_S \lambda_S |S| = \frac{v(N)}{|N|}|N| = v(N)$.

Hence, $\sum_S \lambda_S v(S) \leq v(N)$.

\subsection{Super-additivity}

$\textbf{For all non-empty} \ \mathbf{S, T \in N} \ \textbf{with} \ \mathbf{S \cap T = \phi,} \\ \mathbf{v(S \cup T) \geq v(S) + v(T).}$\vspace{0.1in}

Since $S \subset S \cup T$, from the monotonicity property, we have,

\begin{equation*}
\sum\limits_{j=1}^{M} \left((1-H(|\underset{\forall i \in S \cup T}{max}\left(p_{ij}.d_j\right)|))/c_{S \cup T}(j)\right) \geq
\end{equation*}

\begin{equation}
\sum\limits_{j=1}^{M} \left((1-H(|\underset{\forall i \in S}{max}\left(p_{ij}.d_j\right)|))/c_S(j)\right).
\label{6}
\end{equation}

Similarly, since $T \subset S \cup T$, from the monotonicity property, we have,

\begin{equation*}
\sum\limits_{j=1}^{M} \left((1-H(|\underset{\forall i \in S \cup T}{max}\left(p_{ij}.d_j\right)|))/c_{S \cup T}(j)\right) \geq
\end{equation*}

\begin{equation}
\sum\limits_{j = 1}^{M} \left((1-H(|\underset{\forall i \in T}{max}\left(p_{ij}.d_j\right)|))/c_T(j)\right).
\label{7}
\end{equation}

Multiplying \eqref{6} by $|S|$ and \eqref{7} by $|T|$ and adding together, we have,

\begin{equation*}
|S|\sum\limits_{j=1}^{M} \left((1-H(|\underset{\forall i \in S \cup T}{max}\left(p_{ij}.d_j\right)|))/c_{S \cup T}(j)\right) +
\end{equation*}

\begin{equation*}
|T|\sum\limits_{j=1}^{M} \left((1-H(|\underset{\forall i \in S \cup T}{max}\left(p_{ij}.d_j\right)|))/c_{S \cup T}(j)\right) \geq
\end{equation*}

\begin{equation*}
|S|\sum\limits_{j=1}^{M} \left((1-H(|\underset{\forall i \in S}{max}\left(p_{ij}.d_j\right)|))/c_S(j)\right) +
\end{equation*}

\begin{equation}
|T|\sum\limits_{j=1}^{M} \left((1-H(|\underset{\forall i \in T}{max}\left(p_{ij}.d_j\right)|))/c_T(j)\right).
\label{8}
\end{equation}

Thus,
\begin{equation*}
|S \cup T|\sum\limits_{j=1}^{M} \left((1-H(|\underset{\forall i \in S \cup T}{max}\left(p_{ij}.d_j\right)|))/c_{S \cup T}(j)\right) \geq
\end{equation*}

\begin{equation*}
|S|\sum\limits_{j=1}^{M} \left((1-H(|\underset{\forall i \in S}{max}\left(p_{ij}.d_j\right)|))/c_S(j)\right) +
\end{equation*}

\begin{equation}
|T|\sum\limits_{j=1}^{M} \left((1-H(|\underset{\forall i \in T}{max}\left(p_{ij}.d_j\right)|))/c_T(j)\right).
\label{9}
\end{equation}

Hence, $v(S \cup T) \geq v(S) + v(T)$.

\bibliographystyle{IEEEtran}
\bibliography{Ref}

\end{document}